\documentclass[seceq]{ptptex}

\usepackage{graphicx}

\title{Hadrons in Medium -- Theory confronts experiment}

\author{Fabian \textsc{Eichstaedt}, Stefan \textsc{Leupold}, Ulrich \textsc{Mosel}\footnote{
Speaker, e-mail address: mosel@physik.uni-giessen.de} and Pascal
\textsc{Muehlich}}

\inst{Institut fuer Theoretische Physik, Universitaet Giessen, Giessen,
Germany }

\abst{In this talk we briefly summarize our theoretical
understanding of in-medium selfenergies of hadrons. With the special
case of the $\omega$ meson we demonstrate that earlier calculations
that predicted a significant lowering of the mass in medium are
based on an incorrect treatment of the model Lagrangian; more
consistent calculations lead to a significant broadening, but hardly
any mass shift. We stress that the experimental reconstruction of
hadron spectral functions from measured decay products always
requires knowledge of the decay branching ratios which may also be
strongly mass-dependent. It also requires a quantitatively reliable
treatment of final state interactions which has to be part of any
reliable theory.}

\begin{document}

\maketitle

\section{Introduction}
The study of in-medium properties of hadrons has attracted quite some
interest among experimentalists and theorists alike because of a possible
connection with chiral symmetry restoration in hot and/or dense matter.
Experiments using ultrarelativistic heavy ions reach not only very high
densities, but connected with that also very high temperatures. In their
dynamical evolution they run through various -- physically quite different
-- states, from an initial high-nonequilibrium stage through a very hot
stage of -- possibly - a new state of matter (QGP) to an equilibrated
'classical'  hadronic stage at moderate densities and temperatures. Any
observed signal necessarily represents a time-integral over all these
physically quite distinct states of matter. On the contrary, in
experiments with microscopic probes on cold nuclei one tests interactions
with nuclear matter in a well-known state, close to cold equilibrium. Even
though the density probed is always smaller than the nuclear saturation
density, the expected signals are as large as those from ultrarelativistic
heavy-ion collisions\cite{Mosel_Hirschegg,Mosel:2000fz}.

In this talk we discuss as an example the theoretical situation concerning
the $\omega$ meson in medium and use it to point out various essential
points both in the theoretical framework as well as in the interpretation
of data (for further refs see the reviews in
\cite{Mosel_Erice1,Mosel_Erice2,Rapp-Wambach}).

\section{In-medium Properties: Theory}

The interest in in-medium properties arose suddenly in the early
90's when several authors \cite{Brown-Rho,Hatsuda-Lee} predicted a
close connection between in-medium masses and chiral symmetry
restoration in hot and/or dense matter. This seemed to establish a
direct link between nuclear properties on one hand and QCD
symmetries on the other. Later on it was realized that the
connection between the chiral condensates of QCD and hadronic
spectral functions is not as direct as originally envisaged. The
only strict connection is given by QCD sum rules which restrict only
an integral over the hadronic spectral function by the values of the
quark and gluon condensates which themselves are known only for the
lowest twist configurations. Indeed a simple, but more realistic
analysis of QCD sum rules showed that these do not make precise
predictions for hadron masses or widths, but can only serve to
constrain hadronic spectral functions
\cite{Leupold0,Leupold1,Leupold2,Leupold3}. Thus hadronic models are
needed for a more specific prediction of hadronic properties in
medium.

For example, in the past a lively discussion has been going on about
a possible mass shift of the $\omega$-meson in a nuclear medium.
While there seems to be a general agreement that the $\omega$
acquires a certain width of the order of 40-60 MeV in the medium,
the mass shift is not so commonly agreed on. While some groups have
predicted a dropping mass \cite{Kl99,Kl97,Re02}, there have also
been suggestions for a rising mass \cite{DM01,PM01,SL06,Zs02} or
even a structure with several peaks \cite{Lu02, Mu06}. In this
context a recent experiment by the CBELSA/TAPS collaboration is of
particular interest, since it is the first indication of a downward
shift of the mass of the $\omega$-meson in a nuclear medium
\cite{Tr05}. Since Klingl et al. \cite{Kl97} were among the first to
predict such a downward shift it is worthwhile to look into their
approach again.

The central quantity that contains all the information about the
properties of an $\omega$ meson in medium is the spectral function
\begin{equation}
A_{med}(q) = -\frac{1}{\pi}
\text{Im}\frac{1}{q^2 - (m_\omega^0)^2- \Pi_{vac}(q) - \Pi_{med}(q)}, \label{spectral-function}
\end{equation}
with the bare mass $m_\omega^0$ of the $\omega$. The vacuum part of
the $\omega$ selfenergy $\Pi_{vac}$ is dominated by the decay
$\omega \rightarrow \pi^+ \pi^0 \pi^-$ \cite{Kl96}. For the
calculation of the in-medium part one can employ the
low-density-theorem \cite{Kl97,Lu02,Mu06} which states that at
sufficiently small density of the nuclear medium one can expand the
selfenergy in orders of the density $\rho$
\begin{equation}
\Pi_{med}(\nu, \vec q = 0; \rho) = - \rho T(\nu) \label{low-density-theorem}\ ,
\end{equation}
where $T(\nu)$ is the $\omega$-nucleon forward-scattering amplitude. We
note that a priori it is not clear up to which densities this
low-density-theorem is reliable \cite{Post:2003hu}.

To obtain the imaginary part of the forward scattering amplitude via
Cutkosky's Cutting Rules Klingl et. al \cite{Kl99,Kl97} used an
effective Lagrangian that combined chiral SU(3) dynamics with VMD.
The $\omega$ selfenergy was evaluated on tree-level which needs as
input the inelastic reactions $\omega N \rightarrow \pi N$ ($1\pi$
channel) and $\omega N \rightarrow 2\pi N$ ($2\pi$ channel) to
determine the effective coupling constants. The amplitude $\omega N
\rightarrow \pi N$ is more or less fixed by the measurable and
measured back reaction \cite{Fr97}. This is in contrast to the
reaction $\omega N \leftrightarrow \rho N$ which -- in the
calculations of ref.\cite{Kl99,Kl97} -- is not constrained by any
data and which dominates the $2\pi$ channel. Furthermore, Klingl et.
al \cite{Kl99,Kl97} employed a heavy baryon approximation (HBA) to
drop some of the tree-level diagrams generated by their Lagrangian.
All the calculations were made for isospin-symmetric nuclear matter
at temperature $T=0$. The scattered $\omega$ was taken to have $\vec
q = 0$ relative to the nuclear medium.

We have repeated these calculations without, however, invoking the
HBA.\footnote{For further details of the present calculations we
refer to ref.\cite{Ei06}.} For the $2\pi$ channel which decides
about the in-medium mass shift of the $\omega$ in the calculations
of ref. \cite{Kl99,Kl97} we find considerable differences -- up to
one order of magnitude in the imaginary part of the selfenergy --
when comparing calculations using the full model with those using
the HBA \cite{Ei06}. We thus have to conclude already at this point
that the HBA is unjustified for the processes considered here and
leads to grossly incorrect results.

\begin{figure}
\centerline{\includegraphics[width=8cm]{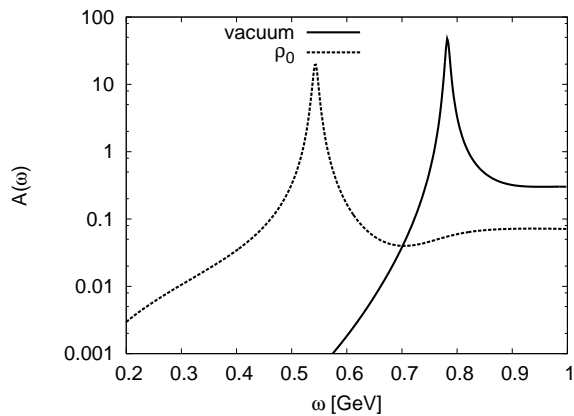}}
\caption{Spectral function of the $\omega$ meson in the vacuum and
at normal nuclear density.} \label{omega-spec-func}
\end{figure}

We show our resulting in-medium spectral function of the $\omega$ (where
HBA was not employed) in figure \ref{omega-spec-func}.  Note that in the
medium the peak is shifted to 544 MeV which is due to the large effects of
a relativistic, full treatment of the imaginary and real parts of the
amplitudes obtained in the present model. This has to be compared with the
results obtained by Klingl et al. \cite{Kl99}. Since Klingl et al. find an
in-medium peak at about 620 MeV it is obvious that in the relativistic
calculation the physical picture changes drastically.
It is also obvious that the correct treatment of the same Lagrangian
as used in ref.\cite{Kl99} on tree-level leads to an unrealistic
lowering of the $\omega$ spectral function.

\begin{figure}
\centerline{\includegraphics[width=8.7cm]{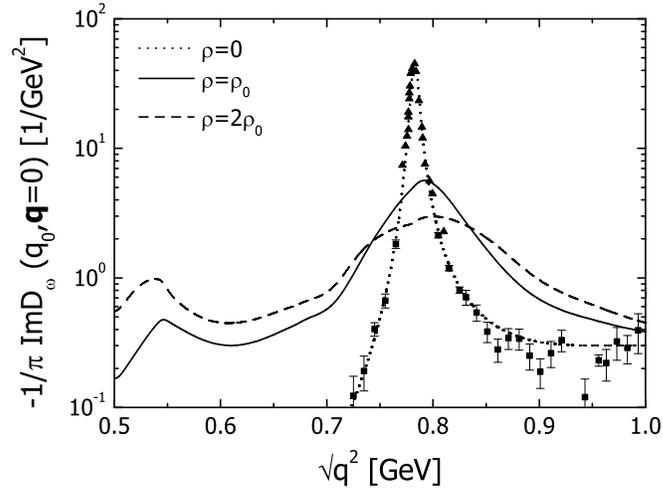}}
\caption{$\omega$ spectral function for an $\omega$ meson at rest,
i.e.~$q_0=\sqrt{q^2}$  (from ref.\ \cite{Mu06}). The appropriately
normalized data points correspond to the reaction
$e^+e^-\rightarrow\omega\rightarrow 3\pi$ in vacuum. Shown are
results for densities $\rho=0$, $\rho=\rho_0=0.16~\mathrm{fm}^{-3}$
(solid) and $\rho=2\rho_0$ (dashed).}\label{Komega}
\end{figure}

It is, therefore, worthwhile to look into another method to
calculate the $\omega$ selfenergy that takes experimental
constraints as much as possible into account and -- in contrast to
the tree-level calculations of ref. \cite{Kl99} -- respects
unitarity. A first study in this direction has been performed by
Lutz et al. \cite{Lu02} who solved the Bethe-Salpeter equation with
local interaction kernels. These authors found a rather complex
spectral function with a second peak at lower energies due to a
coupling to nucleon resonances with masses of about $\approx 1500$
MeV. We have recently used a large-scale K-matrix analysis of all
available $\gamma N$ and $\pi N$ data
\cite{Penner:2002ma,Penner:2002md,Shklyar:2006xw} that does respect
unitarity and thus constrains the essential $2\pi$ channel by the
inelasticities in the $1 \pi$ channel \cite{Mu06}. By consistently
using the low-density-approximation we have obtained the result
shown in Fig. \ref{Komega}.
Fig. \ref{Komega} clearly exhibits a broadened $\omega$ spectral
function with only a small (upwards) shift of the peak mass. In
agreement with the calculations of Lutz et al. \cite{Lu02}, although
with less strength, it also exhibits a second peak at masses around
550 MeV that is due to a coupling to a N*(1535)-nucleon hole
configuration. Such a resonance-hole coupling is known to play also
a major role in the determination of the $\rho$ meson spectral
function \cite{Peters:1997va,Post:2003hu}; in the context of QCD sum
rules it has been examined in ref.\cite{SL06}. It is obviously quite
sensitive to the detailed coupling strength of this resonance to the
$\omega N$ channel which energetically opens up only at much higher
masses.

\begin{figure}
\centerline{\includegraphics[width=8cm]{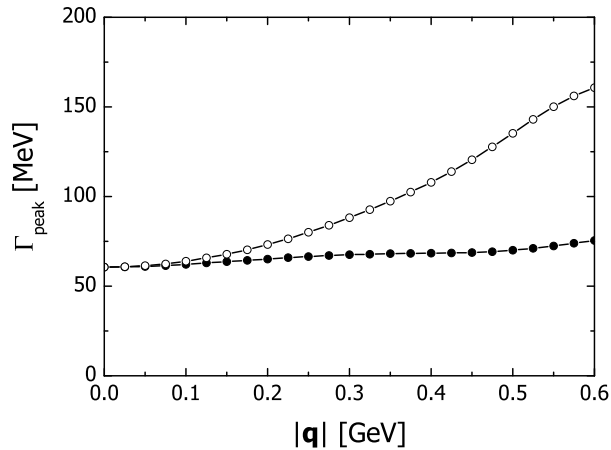}}
\caption{On-shell width of the $\omega$ in nuclear matter at nuclear
matter density $\rho_0$  (from ref.\ \cite{Mu06}). The open (solid)
points give the width for the transverse (longitudinal) degree of
freedom.} \label{Womega}
\end{figure}

As mentioned earlier, there is general consensus among different theories,
that  the on-shell width of the $\omega$ meson in medium reaches values of
about 50 MeV at saturation density. To illustrate this point we show in
Fig. \ref{Womega} the width as a function of omega momentum relative to
the nuclear matter restframe both for the transverse and the longitudinal
polarization degree of freedom.
It is clearly seen that the transverse width increases strongly as a
function of momentum. At values of about 500 MeV, i.e. the region, where
CBELSA/TAPS measures, the transverse width has already increased to about
125 MeV and even the polarization averaged width amounts to 100 MeV.

\section{Spectral Functions and Observables}

\begin{figure}[h]
\centerline{\includegraphics[width=8cm]{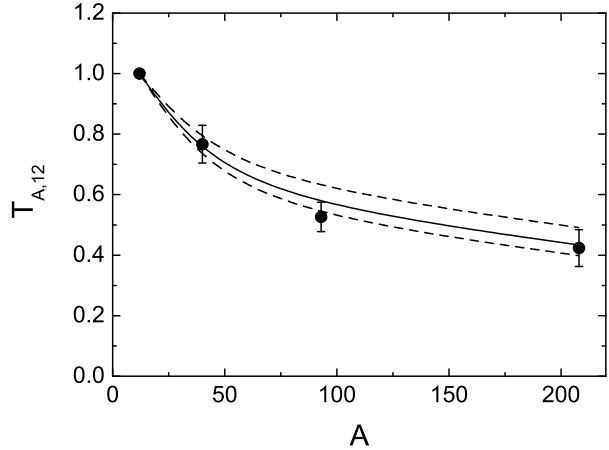}}
\caption{Transparency of nuclei for $\omega$ production.
Calculations and preliminary data are normalized to $^{12}C$. Dashed
lines reflect error estimates obtained from the spread of the data.
Data are from CBELSA/TAPS\cite{Kot}.} \label{Tomega}
\end{figure}

Apart from invariant mass measurements, there is another possibility
to experimentally constrain the in-medium broadening of the
$\omega$-meson. The total width plotted in Fig. \ref{Womega} is the
sum of elastic and inelastic widths. In general, the inelastic width
alone is determined by the imaginary part of the selfenergy and the
latter determines the amount of reabsorption of $\omega$ mesons in
the medium. In a Glauber approximation the cross section for
$\omega$ production on a nucleus reads
\begin{equation}
\frac{d\sigma_{\gamma + A \to \omega + X}}{d\Omega} = \int d^3x\,
\rho(\vec{x}) \frac{d\sigma_{\gamma + N \to \omega + X}}{d\Omega}
\exp\left[-\int_z^\infty dz'\,\left(- \frac{1}{p} \Im
\Pi(p,\rho(\vec{x}\,'))\right)\right]~.
\end{equation}
The ratio of this cross section on the nucleus to that on the
nucleon then determines the nuclear transmission $T$ which depends
on the imaginary part of the omega selfenergy $\Im \Pi$
\begin{equation}
T(A) \approx \int d^3x\, \rho(\vec{x}) \exp\left[-\int_z^\infty
dz'\, \left(- \frac{1}{p} \Im
\Pi(p,\rho(\vec{x}\,'))\right)\right]~.
\end{equation}
Using in addition the low-density-approximation
\begin{equation}    \label{LDAPP}
\Im \Pi(p,\vec{x}) = - p \rho(\vec{x}) \sigma_{\omega N}^{\rm
inel} ~
\end{equation}
one obtains the usual Glauber result
\begin{equation}
T(A) = \int d^3x\, \rho(\vec{x}) \exp\left[-\int_z^\infty dz'\,
\rho(\vec{x}\,') \sigma_{\omega N}^{\rm inel} \right]~.
\end{equation}
We show the calculated transmission $T$ in Fig. \ref{Tomega}
together with the data obtained by CBELSA/TAPS.
The measured cross section dependence on massnumber $A$ is
reproduced very well \cite{Tomega} if the inelastic $\omega N$ cross
section is increased by 25\% over the usually used parametrization.
This may indicate a problem with the usually used cross section, or
- more interesting - it may indicate a breakdown of the
low-density-approximation.

It is, furthermore, important to realize that the spectral functions
themselves are not observable. What can be observed are the decay
products of the meson under study. It is thus obvious that even in
vacuum the invariant mass distribution of the decaying resonance ($V
\to X_1 X_2$), reconstructed from the four-momenta of the decay
products ($X_1,X_2$), involves a product of spectral function and
partial decay width into the channel being studied
\begin{equation}
\frac{dR_{V \to X_1X_2}}{dq^2} \sim A(q^2) \times \frac{\Gamma_{V \to X_1
X_2}(q^2)}{\Gamma_{\rm tot}(q^2)}~.
\end{equation}
Since in general the branching ratio also depends on the invariant mass of
the decaying resonance this dependence may distort the observed invariant
mass distribution compared with the spectral function itself. This effect
is obviously the more important the broader the decaying resonance is and
the stronger the widths depend on $q^2$.

While these branching ratios are usually well known in vacuum there is
considerable uncertainty about their value in the nuclear medium. This
uncertainty is connected with the lack of knowledge about the in-medium
vertex corrections, i.e. the change of coupling constants with density.
Even if we assume that these quantities stay the same, then at least the
total width appearing in the denominator of the branching ratio has to be
changed, consistent with the change of the width in the spectral function.
This point has only rarely been discussed so far, but it has far-reaching
consequences.

For example, for the $\rho$ meson the partial decay width into the
dilepton channel goes like
\begin{equation}
\Gamma_{\rho \to e^+e^-} \sim \frac{1}{M^4} M = \frac{1}{M^3} ~,
\end{equation}
where the first factor on the rhs originates in the photon propagator and
the last factor $M$ comes from phase-space. On the other hand, the total
decay width of the $\rho$ meson in vacuum is given by (neglecting the pion
masses for simplicity)
\begin{equation}
\Gamma_{\rm tot} \approx \Gamma_{\rho \to \pi \pi} \sim M ~,
\end{equation}
so that the branching ratio in vacuum goes like
\begin{equation}
\frac{\Gamma_{\rho \to e^+e^-}}{\Gamma_{\rm tot}} \sim \frac{1}{M^4}~.
\end{equation}
This strong $M$-dependence distorts the spectral function, in particular,
for a broad resonance such as the $\rho$ meson. This effect is contained
and clearly seen in theoretical simulations of the total dilepton yield
from nuclear reactions (see, e.~g., Figs.~$8-10$ in
\cite{Effenberger:1999ay}); it leads to a considerable shift of strength
in the dilepton spectrum towards lower masses.

For the semileptonic decay channel $\pi^0 \gamma$ that has been
exploited in the CBELSA TAPS experiment again a strong
mass-dependence of the branching ratio shows up because just at the
resonance the decay channel $\omega \to \rho \pi$ opens up.

In both of these cases the in-medium broadening changes the total
widths in the denominator of the branching ratios even if the
partial decay width stays the same as in vacuum. Such an in-medium
broadening of the total width, which should be the same as in the
spectral function, will tend to weaken the $M$-dependence of the
total width and thus the branching ratio as a whole. In medium
another complication arises: the spectral function no longer depends
on the invariant mass alone, but -- due to a breaking of
Lorentz-invariance because of the presence of the nuclear medium --
in addition also on the three-momentum of the hadron being probed.
Again, this $p$-dependence of the vector meson selfenergy has only
rarely been taken into account (see, however,
refs.\cite{Peters:1997va,Post:2000qi,Mu06}). In addition, final
state interactions do affect hadronic decay channels. A
quantitatively reliable treatment of these FSI thus has to be
integral part of any trustable theory that aims at describing these
data.

\section{Conclusions} \label{conclusion}
QCD sum rules establish a very useful link between the chiral condensates,
both in vacuum and in medium, but their connection to hadronic spectral
functions is indirect. The latter can thus only be constrained by the
QCDSR, but not be fixed; for a detailed determination hadronic models are
needed. We have pointed out in this talk that the
low-density-approximation nearly always used in these studies does not
answer the question up to which densities it is applicable. First
studies~\cite{Post:2003hu} have shown that this may be different from particle to
particle.

While the in-medium properties of all vector mesons $\rho$,
$\omega$, and $\phi$ are the subject of intensive experimental and
theoretical research, in this talk we have concentrated on the
$\omega$ meson for which recent experiments indicate a lowering of
the mass by about 60 MeV in photon-produced experiments on nuclei. A
tree-level calculation, based on an effective Lagrangian, that
predicted such a lowering, has been shown to be incorrect because of
the heavy-baryon approximation used in that calculation. A correct
tree-level calculation with the same Lagrangian gives strong
contributions from the $\omega \to 2\pi N$ channel, which, however,
is unconstrained by any data; in effect, the spectral function is
softened by an unreasonable pole mass shift. This problem might
partially be based on the fact that all the inelastic processes
$\omega N \rightarrow \pi N$ and $\omega N \rightarrow 2\pi N$ are
only treated at tree-level. Here an improved calculation is needed,
which incorporates coupled-channels and rescattering, e.g. a
Bethe-Salpeter \cite{Lu02} or a K-matrix approach
\cite{Penner:2002ma,Fe98,Sh04}.

We have indeed shown that a better calculation that again starts out
from an effective Lagrangian and takes unitarity, channel-coupling
and rescattering into account yields a significantly different
in-medium spectral function in which the pole mass hardly changes,
but a broadening of about 60 MeV at nuclear saturation density takes
place, which increases with momentum, primarily in the transverse
channel.

Finally, we have pointed out that any measurement of the spectral
function necessarily involves also a branching ratio into the
channel being studied. The experimental in-medium signal thus
contains changes of both the spectral function and the branching
ratio.

\section*{Acknowledgements}
The authors acknowledge discussions with Norbert Kaiser and Wolfram
Weise. They have also benefitted a lot from discussions with Vitaly
Shklyar. This work has been supported by DFG through the SFB/TR16
"Subnuclear Structure of Matter".


\end{document}